\documentclass[a4paper,UKenglish,cleveref, autoref, thm-restate]{lipics-v2021}
\nolinenumbers

\usepackage{xspace}
\usepackage{relsize}
\usepackage{multirow}
\usepackage{multicol}
\usepackage{mathtools}

\usepackage{xcolor}
\usepackage{listings}

\newcommand*{\X}{\ensuremath{\mathcal{X}}\xspace}
\newcommand*{\id}{\ensuremath{\mathsf{id}}\xspace}
\newcommand*{\I}{\ensuremath{\mathsf{PK}}\xspace}
\newcommand*{\feat}{\ensuremath{\mathcal{F}}\xspace}

\newcommand*{\C}{\ensuremath{\mathcal{C}}\xspace}

\newcommand*{\V}{\ensuremath{\Sigma^*}\xspace}

\renewcommand{\_}{\textscale{.55}{\textunderscore}}

\title{Global Benchmark Database}
\titlerunning{GBD} %TODO optional, please use if title is longer than one line

\author{Ashlin Iser}{Karlsruhe Institute of Technology, Germany \and \url{https://ae.iti.kit.edu/3986.php}}{ashlin.iser@kit.edu}{https://orcid.org/0000-0003-2904-232X}{}%{(Optional) author-specific funding acknowledgements}%TODO mandatory, please use full name; only 1 author per \author macro; first two parameters are mandatory, other parameters can be empty. Please provide at least the name of the affiliation and the country. The full address is optional. Use additional curly braces to indicate the correct name splitting when the last name consists of multiple name parts.
\author{Christoph Jabs}{HIIT, University of Helsinki, Finland \and \url{https://christophjabs.info}}{christoph.jabs@helsinki.fi}{https://orcid.org/0000-0003-3532-696X}{Work financially supported by Research Council of Finland grant 356046.}

\authorrunning{A.~Iser, C.~Jabs} %TODO mandatory. First: Use abbreviated first/middle names. Second (only in severe cases): Use first author plus 'et al.'

\Copyright{Ashlin Iser, Christoph Jabs} %TODO mandatory, please use full first names. LIPIcs license is "CC-BY";  http://creativecommons.org/licenses/by/3.0/
        
 %TODO mandatory; please add comma-separated list of keywords

\category{Tool Paper} %optional, e.g. invited paper

\relatedversion{} %optional, e.g. full version hosted on arXiv, HAL, or other respository/website
\acknowledgements{Many thanks to all the people who have supported GBD, either by contributing code or data, or by providing infrastructure.
Special thanks also go to the early adopters whose successes and feedback have been and continue to be inspiring and motivating.} 

\EventEditors{Supratik Chakraborty and Jie-Hong Roland Jiang}
\EventNoEds{2}
\EventLongTitle{27th International Conference on Theory and Applications of Satisfiability Testing (SAT 2024)}
\EventShortTitle{SAT 2024}
\EventAcronym{SAT}
\EventYear{2024}
\EventDate{August 21--24, 2024}
\EventLocation{Pune, India}
\EventLogo{}
\SeriesVolume{305}
\ArticleNo{18}

\begin{document}

\maketitle

\begin{abstract}
This paper presents Global Benchmark Database (GBD),
a comprehensive suite of tools for provisioning and sustainably maintaining benchmark instances and their metadata.
The availability of benchmark metadata is essential for many tasks in empirical research, e.g., for the data-driven compilation of benchmarks, the domain-specific analysis of runtime experiments, or the instance-specific selection of solvers.
In this paper, we introduce the data model of GBD as well as its interfaces and provide examples of how to interact with them.
We also demonstrate the integration of custom data sources and explain how to extend GBD with additional problem domains, instance formats and feature extractors.
\end{abstract}

\section{Introduction}
\label{sec:intro}

The idea to create Global Benchmark Database (GBD) arose from the need to make available and sustainably maintain benchmark instances and their metadata for the problem of propositional satisfiability (SAT).
To this end, we specified a simple hash function for \emph{instance identification} to serve as the primary key for SAT instance data sets.
A proof of concept was presented and discussed at the 2018 Pragmatics of SAT (POS) workshop, demonstrating the specification of a SAT instance identifier and methods for labeling and querying for instances~\cite{Iser:2018:GBD}.
From there, GBD has matured into a comprehensive suite of tools for provisioning and sustainably maintaining benchmark instances of various problem domains and their metadata.
Supported problem domains include propositional satisfiability (SAT), maximum satisfiability (MaxSAT), and pseudo-Boolean optimization (PBO).

In essence, the purpose of GBD is to act as a conduit between data science and empirical research on NP-hard problem classes by facilitating the seamless integration of benchmark data into existing workflows.
GBD provides tools for distributing benchmark instances and feature databases, as well as tools for transforming instances and extracting instance features.
It also provides a set of prebuilt feature databases for identifying instance equivalence classes, categories, or labels, as well as analytical instance features.
Examples of successful applications of GBD include the sanitization and selection of benchmarks for SAT~competitions~\cite{Iser:SC2023Bench}, domain-specific solver evaluations~\cite{SC2020:AIJ}, and the analysis of solver portfolios and solver prediction models~\cite{Bach:2022:kPortfolios}.
Moreover, the authors of the latest award-winning state-of-the-art SAT solvers use GBD in their empirical evaluations.~\cite{Haberlandt:2023:SBVA,Schreiber:2023:Diss}.
GBD is open source and available under the MIT license~\cite{Iser_GBD_Tools_2023}.

In this paper, we describe the conceptual design of the GBD data model and present the interfaces for interacting with GBD, as well as the extension capabilities provided by the data model, including concrete examples of their use.
We start with a brief description of the GBD data model and query language in Section~\ref{sec:model}.
Section~\ref{sec:applications} presents the GBD interfaces and gives concrete examples of how to use GBD in practice.
Finally, Section~\ref{sec:extensions} explains how to extend GBD with additional problem domains, instance formats, and feature extractors.
We conclude with a summary and an outlook on future work in Section~\ref{sec:outlook}.

\section{Data Model and Queries}\label{sec:model}

Conceptually, GBD consists of an extensible set of contexts $\C = \{C_0, C_1, \dots\}$.
Each context represents a problem domain, or more specifically, a particular instance format of a problem domain.
A context is defined as a tuple $C = (\X, \id)$, where $\X$ is the set of benchmark instances and $\id:\X \rightarrow \I$ is a context-specific instance identification function that maps each benchmark instance to an instance identifier unique within the context.
% NOTE: technically need to define $\V$ first
Contexts are instantiated with data sources $D = \{ \feat_1, \feat_2, \dots \}$ which provide instance features $\feat_i \subseteq \I \times 2^{\V}$ assigning values to benchmark instances, where values are represented by strings over an alphabet $\Sigma$.\footnote{The data sources used in the examples of this section are available at \url{benchmark-database.de}.}

\paragraph*{Feature Types.}
In practice, GBD distinguishes between one-to-one and one-to-many features.
One-to-one features are defined as those that provide a single value per instance, with a default value associated with them.
In contrast, one-to-many features can have zero to any number of values per instance.
One-to-one features facilitate the initial setup and subsequent maintenance, as they are initialized to the default value automatically and subsequent calls to \textsf{gbd set} overwrite any previous settings.
In contrast, the values of one-to-many features accumulate over time and must be explicitly cleared.
Example~\ref{ex:features} illustrates the practical use of the different feature types in the context of \textsf{cnf} instances.

\begin{example}\label{ex:features}
Let $\textsf{cnf}:=(\X, \id)$ be the context where \X is the set of SAT benchmark instances in DIMACS CNF format, and let a data source $\textsf{meta.db} := \{ \mathsf{track}, \mathsf{result}, \dots \}$ be given, which provides the \textsf{cnf} instance features \textsf{track} and \textsf{result}.
In the example, the feature \textsf{track} indicates the competition tracks in which the instance was used and is modeled as a one-to-many feature, while \textsf{result} indicates the satisfiability of the instance and is defined as a one-to-one feature with the default value ``unknown''.
\end{example}

\paragraph*{Context Mappings.}
For any two contexts $C_i = (\X_i, \id_i)$, $C_j = (\X_j, \id_j)$ with instance identifiers $\I_i, \I_j$, a data source can provide a special feature $\feat \subseteq \I_i \times \I_j$ that establishes a relationship between instances in $\X_i$ and $\X_j$.
In GBD, such features are referred to as context mappings.
Context mappings can be used to map instances from different problem domains to each other if there is a reduction procedure that transforms instances from one domain to instances in the other domain.
Example~\ref{ex:map} illustrates the practical use of context mappings to relate instances of the propositional satisfiability problem to instances of the $k$-independent set problem.

\begin{example}\label{ex:map}
Given the context of $\textsf{cnf}$ instances of the SAT problem and the context $\textsf{kis}$ of graph-instances of the $k$-Independent~Set problem.
Any SAT instance can be transformed to a graph and a number $k$ in such a way that the satisfiability of the SAT instance is equivalent to the existence of an independent set of size $k$ in the graph~\cite{Papadimitriou:1994:Complexity}. 
A context mapping feature can be used to map instances in $\textsf{cnf}$ to their corresponding instances in $\textsf{kis}$. 
\end{example}%

\noindent
Context mappings are useful not only for relating instances from different problem domains, but also for relating instances from different contexts within the same problem domain.
Application scenarios include relating different instance formats, analyzing different encodings of application instances, or distinguishing sanitized or otherwise preprocessed instances from their original counterparts (cf.\ Use case~\ref{cmd:gbd:sanitize} below).

\paragraph*{Queries.}
GBD provides a query language designed to filter instances of a given context by specifying constraints over the instance features.
A GBD query is a simple constraint $c$, or a compound constraint $c_1~\texttt{and}~c_2$, or $c_1~\texttt{or}~c_2$ for sub-constraints $c_1$, $c_2$, and parentheses can be used to indicate precedence.
A simple constraint is of the form $f \circ e$ with a feature name $f$, a value $e$, and an operator $\circ \in \{ \texttt{=, !=, <, >, <=, >=, like, unlike} \}$ with the usual semantics. 
For SQL-inspired operators $\circ \in \{ \texttt{like, unlike} \}$, a constraint can also be of the form $f \circ \texttt\%e$, $f \circ e\texttt\%$, or $f \circ \texttt\%e\texttt\%$ to indicate postfix, prefix, or infix constraints, respectively.
Value $e$ can also be a term, where a term is either a constant number, or a feature name in parentheses $(f')$\footnote{Parentheses are used here to disambiguate terms and character strings.}, or a compound term $(t_1 \circ t_2)$ from arithmetic operators $\circ \in \{ \texttt{+,-,*,/} \}$ and terms $t_1$, $t_2$. 

\begin{example}\label{ex:query}
    Given the \textsf{cnf} context and the data source \textsf{meta.db} providing the instance feature \textsf{track}, a query can be of the form $\mathsf{track} = \texttt{main\_2023}$ to filter for instances from the Main track of the 2023 SAT competition.
\end{example}

\begin{example}\label{ex:query2}
    Given the \textsf{cnf} context and the data source \textsf{meta.db} providing the instance features \textsf{track} and \textsf{filename}, a query can be of the form $\mathsf{track} = \texttt{anni\_2022} \textsf{ and } \mathsf{filename} \textsf{ like } \texttt{rphp\%}$ to filter for instances from the Anniversary track of the 2022 SAT competition whose filenames begin with the string \texttt{rphp}.
\end{example}

\begin{example}\label{ex:query3}
    Given the \textsf{cnf} context and the data source \textsf{base.db} providing the instance features \textsf{variables} and \textsf{clauses}, a query can be of the form $\mathsf{variables} > \texttt{(clauses)}$ to filter for instances with more variables than clauses. 
\end{example}

\section{Applications}\label{sec:applications}

This section introduces the \textsf{gbd-tools} package, which is available in the Python Package Index (PyPI) and can be installed using the command \textsf{pip install gbd-tools}. 
The package provides the command line tool \textsf{gbd}, a web service, and the Python interface class \textsf{GBD}.
We start with instructions on how to configure \textsf{gbd-tools} in Section~\ref{sec:config}, and then present use cases of the command line tool \textsf{gbd} in Section~\ref{sec:cli}.
The web service and Python interface are presented in Sections~\ref{sec:web} and~\ref{sec:python}, respectively.

\subsection{Data Source Configuration}\label{sec:config}
Data sources for instantiating GBD are specified as a list of files, where each file can be either a \textsf{sqlite3} database created by GBD, or a \textsf{csv} file for importing data from other sources.
The requirements for the \textsf{csv} files are that they have a header line containing the feature names, and they must provide the \textsf{hash} column containing the instance identifiers.
Example~\ref{ex:datasource} shows a typical example of integrating two data sources before analyzing runtime experiments.
\begin{example}\label{ex:datasource}
Let GBD be instantiated with the two data sources \textsf{meta.db} and \textsf{runtimes.csv}.
The file \textsf{meta.db} provides the feature \textsf{family}, indicating the domain of each instance.
The file \textsf{runtimes.csv} provides the solver runtimes \textsf{baseline} and \textsf{incumbent}.
The instance identifier in the \textsf{hash} column facilitates the integration of the two data sources and thus the domain-wise analysis of experimental results.
\end{example}%
\noindent
Data sources can be registered by setting the environment variable \textsf{GBD\_DB} to a colon-separated list of paths.
It is also possible to specify or override the data sources with the option \textsf{-d/-{-}db} on each call of the \textsf{gbd} command-line tool.
If a data source does not exist, GBD offers to create it.
Ready-made feature databases are available for download from \url{https://benchmark-database.de}.
The \textsf{gbd info} command can be used to display the registered data sources, their names, and the features provided.
The name of a data source is automatically generated from its filename by removing the extension, and is needed in some parameters and queries to explicitly refer to a specific data source (cf.~Use cases~\ref{cmd:gbd:explicitdb} and~\ref{cmd:gbd:init:base}).

Each data source is associated with a context, so a data source can only provide features for exactly one context.
Each context is identified by a name, e.g.,~\textsf{opb} for the context of pseudo-Boolean optimization instances.
The set of available contexts, including their names and descriptions, can be displayed with the command \textsf{gbd info -c}.
To bind a data source to a context, its filename is prefixed with the context name to which the data belongs (cf.~Use cases~\ref{cmd:gbd:sanitize} and~\ref{cmd:gbd:transform}).
Otherwise, the data source is treated as belonging to the default context \textsf{cnf}.

When features from different contexts are queried simultaneously, a context mapping feature is used to create a relational join between the instance identifiers of the different contexts.
The naming convention for context mapping features is \textsf{to\_\{cxt\}}, where \textsf{cxt} is a context name.
The \textsf{to\_\{cxt\}} feature maps the identifiers of the context of the data source in which it resides to the identifiers of the context named by \textsf{cxt}.
GBD recognizes context mapping features according to this naming convention and generates the appropriate foreign key relationships to automatically join features from different contexts.

\subsection{Command-Line Tool \textsf{gbd}}\label{sec:cli}

In this section, we present the command line tool \textsf{gbd} and its subcommands for initializing GBD with benchmark instances, querying for instances and features, extracting features from instances, and transforming instances.
The full set of subcommands can be viewed with \textsf{gbd -{-}help}.
In the use cases presented in this section, we assume that the \textsf{cnf} data sources \textsf{meta.db}\footnote{\url{https://benchmark-database.de/getdatabase/meta}} and \textsf{mylocal.db} (to be initialized in the following) are configured as data sources by the environment variable \textsf{GBD\_DB}, and that \textsf{/path/to/instances} contains the benchmark instances from SAT~Competition~2023.\footnote{\url{https://benchmark-database.de/?track=main_2023}}

\subsubsection{Database Initialization}
In order to use GBD to organize your own experiments, it is necessary for GBD to know where the locally available benchmark instances are located.
Integrating a set of paths to benchmark instances into GBD requires computing the instance identifiers and storing the instance paths in a GBD data source.
In this process, GBD creates the reserved features \textsf{local} and \textsf{filename} to associate the local paths and filenames with the respective identifiers.
Note that projecting to an instance identifier automatically eliminates duplicates.
The initialization process may require a considerable amount of time, even for a moderate number of instances, as is the case with SAT competition benchmark sets. However, it is only required once for the registration of benchmarks.

Use case~\ref{cmd:gbd:init} illustrates initializing GBD with a number of benchmark instances present on the local volume using the \textsf{gbd init local} command.
The \textsf{-j/-{-}jobs} option is used to set the number of parallel jobs, the \textsf{-{-}target} option is used to specify the name of the database in which to create the features, and the parameter after the \textsf{local} subcommand is used to specify the path to the benchmark instances.
The \textsf{gbd init local} command recursively searches the specified directory for benchmark instances using the file extensions associated with the context of the target database.
\begin{lstlisting}[language=bash,caption={Local database initialization},label={cmd:gbd:init}]
gbd init -j16 --target mylocal local /path/to/instances
\end{lstlisting}%

\subsubsection{Queries for Instances and Features}
The \textsf{gbd get} command is used to filter for specific instances and retrieve their features.
A comprehensive set of options can be displayed with \textsf{gbd get -{-}help}.
Filtering is typically done with GBD queries, and a list of features to be returned is given by the \textsf{-r/-{-}resolve} option.
Use case~\ref{cmd:gbd:get} illustrates how to filter for cryptographic instances in the SAT~Competition~2023 benchmark set and return their \textsf{local} paths.

\begin{lstlisting}[language=bash,caption={Filtering and feature resolution},label={cmd:gbd:get}]
gbd get "track=main_2023 and family like crypto%" -r local
\end{lstlisting}%
\noindent
Feature names from different databases may overlap in some configurations.
In this case, GBD picks the feature values from the first database in the list that provides the feature.
To override this behavior, feature names can be prefixed with the database name, separated by a colon.
Use case~\ref{cmd:gbd:explicitdb} shows how to filter by the feature \textsf{track} from the database \textsf{meta} and select the feature \textsf{local} from the database \textsf{mylocal}.
\begin{lstlisting}[language=bash,caption={Explicit feature database},label={cmd:gbd:explicitdb}]
gbd get "meta:track=main_2023" -r mylocal:local
\end{lstlisting}%
\noindent
To control the handling of multiple values per hash, the \textsf{-c/-{-}collapse} option can be used, as shown in Use case~\ref{cmd:gbd:collapse}.
The command collapses the values of the feature \textsf{local} to a single value by using the minimum value.
\begin{lstlisting}[language=bash,caption={Collapse},label={cmd:gbd:collapse}]
gbd get "track=main_2023" -r mylocal:local -c min
\end{lstlisting}%
\noindent
For grouping instances by a feature other than the instance identifier, the option \textsf{-g/-{-}group} can be used as illustrated in Use case~\ref{cmd:gbd:group}.
The command groups the instances by the feature \textsf{isohash} and selects a local path for each group using the minimum function.
\begin{lstlisting}[language=bash,caption={Grouping},label={cmd:gbd:group}]
gbd get "track=main_2023" -r local -c min -g isohash
\end{lstlisting}%
\subsubsection{Manual Data Acquisition}
In addition to subcommands for creating, deleting, renaming, and copying features, the \textsf{gbd} command line tool provides a subcommand for manually setting feature values.
Use case~\ref{cmd:gbd:set} demonstrates the use of the \textsf{gbd set} command to set the value of the \textsf{family} feature to \textsf{hardware-verification} for all instances whose filename starts with the string \textsf{manol}.
\begin{lstlisting}[language=bash,caption={Setting feature values},label={cmd:gbd:set}]
gbd set family=hardware-verification "filename like manol%"
\end{lstlisting}%

\subsubsection{Instance Feature Extraction}
Feature extractors are accessible through subcommands of \textsf{gbd init}, which can be listed with \textsf{gbd init -{-}help}.
Each feature extractor is bound to a set of contexts and can only be executed on instances from those contexts.
Multiple feature extractors are provided by the \textsf{gbdc} extension module which are documented at \url{{https://udopia.github.io/gbdc}}.
Note that ready-made feature databases are available for download from our instance of the GBD web interface (cf.~Section~\ref{sec:web}).

Use case~\ref{cmd:gbd:init:base} illustrates the use of the subcommand \textsf{gbd init base} to extract default features denoted as \textsf{base} from a set of \textsf{cnf} benchmark instances.
In the example, the option \textsf{-j/-{-}jobs} is used to specify the number of parallel jobs to use for the feature extraction.
The option \textsf{-{-}target} is used to specify the name of the database in which to create the features.
This requires that a database with the name \textsf{mybase} is registered as a data source.
The parameter after the subcommand \textsf{base} is used to specify the query for the instances to extract features from.
\begin{lstlisting}[language=bash,caption={Base feature extraction},label={cmd:gbd:init:base}]
gbd init -j16 --target mybase base "track=main_2023"
\end{lstlisting}%
\noindent
It is often necessary to determine which instances are isomorphic to each other.
The \textsf{isohash} feature is a hash value that over-approximates the class of isomorphic instances and can be used to identify and group isomorphic instances and to subsequently eliminate them from benchmark sets (cf.~Use case~\ref{cmd:gbd:group}).
Use case~\ref{cmd:gbd:init:isohash} illustrates how to use the \textsf{gbd init isohash} command to compute \textsf{isohash} for a set of \textsf{cnf} benchmark instances.
\begin{lstlisting}[language=bash,caption={Isohash calculation},label={cmd:gbd:init:isohash}]
gbd init -j16 --target meta isohash "track=main_2023"
\end{lstlisting}%

\subsubsection{Instance Transformation}
GBD also provides instance transformers accessible via subcommands of \textsf{gbd transform}.
A complete list of available instance transformers can be displayed with \textsf{gbd transform -{-}help}.
Instance transformers are functions that transform an instance of one context into another instance of a different context.
Additionally, an instance transformer automatically creates a context mapping that relates the instances in both contexts, and can store additional features in the database.
Instance transformers are implemented as a special type of feature extractor that, in addition to creating instance features, also creates a new benchmark instance in the target context.
Several instance transformers are provided by the \textsf{gbdc} extension module and are documented in the \textsf{gbdc} documentation.

Use case~\ref{cmd:gbd:sanitize} illustrates the use of the sanitize transformer for \textsf{cnf} instances.
The command transforms the instances in the source context \textsf{cnf} to instances in the target context \textsf{sancnf} and stores their mapping and local path in the database \textsf{sancnf\_local.db}.
\begin{lstlisting}[language=bash,caption={Instance sanitizer},label={cmd:gbd:sanitize}]
gbd transform --source cnf --target sancnf_local sanitize
\end{lstlisting}%
\noindent
Note that \textsf{sancnf} is an example of a context that exists only to distinguish between sanitized and unsanitized instances and to use the context mapping feature to check if identifiers have changed.
In most practical use cases, the sanitized instances will be used to replace the unsanitized instances in the original \textsf{cnf} context.

An example of a real transformation between different problem domains is the transformation of \textsf{cnf} instances into $k$-Independent Set (\textsf{kis}) instances, as described in Example~\ref{ex:map}.
Use case~\ref{cmd:gbd:transform} illustrates the use of the \textsf{cnf2kis} transformer to transform the instances in the source context \textsf{cnf} into instances in the target context \textsf{kis}.
The mapping of the instances in both contexts is stored in the database \textsf{kis\_local.db}.
\begin{lstlisting}[language=bash,caption={Transforming CNF to KIS},label={cmd:gbd:transform}]
gbd transform --source cnf --target kis_local cnf2kis
\end{lstlisting}%

\subsection{Web Interface}\label{sec:web}%
The GBD web interface is implemented as a RESTful web service~\cite{Fielding:2000:REST} that provides access to the benchmarks and databases of the environment it runs in, and can be started with the command \textsf{gbd serve}.
While the GBD command-line tool operates independently of the web service, the web service uses the same GBD API and database configuration as the command-line tool.
Our instance of the GBD web interface, accessible via \url{{https://benchmark-database.de}}, is configured to run behind an \textsf{Nginx} reverse proxy~\cite{Reese:2008:Nginx} and is hosted in a \textsf{Docker} container~\cite{Merkel:2014:Docker}.
At the time of writing, we provide access to more than 100\,000 benchmark instances from the \textsf{cnf}, \textsf{wcnf}, and \textsf{opb} contexts, with prebuilt feature databases for them.

\subsection{Python Interface}\label{sec:python}
The GBD Python API is wrapped in the class \textsf{gbd\_core.api.GBD} and its documentation can be found at \mbox{\url{https://udopia.github.io/gbd}}.
With the Python API, GBD data sources can be directly integrated into Python scripts and Jupyter notebooks to be used in evaluation and analysis.
Use case~\ref{cmd:gbd:api} illustrates the use of the Python API to query for instances and features.
Line 2 creates a GBD object that connects to the GBD data sources \textsf{base} and \textsf{meta}, which are passed to the constructor as a list of paths.
Line 3 creates a list of feature names from the \textsf{base} data source and adds the feature \textsf{family} from the \textsf{meta} data source.
Line 4 queries for the given features of \textsf{cnf} instances from the Main~track of SAT~Competition~2023.
The result is returned as a Pandas~\cite{McKinney:2010:Pandas} DataFrame for further analysis.\footnote{We provide illustrative examples of benchmark data analyses, including portfolio analysis, category prediction, and category-specific ranking, accessible via \url{https://udopia.github.io/gbdeval/}.}
\begin{lstlisting}[language=python,caption={GBD Python interface},label={cmd:gbd:api},numbers=left,stepnumber=1,numberstyle=\tiny,numbersep=5pt]
from gbd_core.api import GBD
with GBD(['gbd/meta.db', 'gbd/base.db']) as gbd:
    feat = gbd.get_features('base') + ['family']
    df = gbd.query("track=main_2023", resolve=feat)
\end{lstlisting}%

\section{Writing Extensions}\label{sec:extensions}

The architectural design of GBD allows the integration of new contexts, feature extractors, and instance transformers.
The extensibility of the system is achieved through the use of dictionaries, which serve as registries for contexts, feature extractors, and instance transformers.
The following section describes the elements required to create such a registry entry. 
It is important to note that at the time of writing, these registries are hard-coded, which is a limitation in that it requires modification of these dictionaries within the source code of GBD.
This issue will be addressed in future versions of GBD, where the registry will be moved to configuration files.

GBD contexts are managed in a dictionary data structure that is initialized in the \textsf{gbd\_core.contexts} module.
The dictionary is indexed with the respective \emph{context name} and contains a pointer to the instance \emph{identification function} and a list of valid instance \emph{filename extensions}.
The instance identification function is the function that assigns a unique instance identifier to each benchmark instance, which serves as the primary key in the context-specific feature databases.
The list of valid instance filename extensions is used to locate benchmark instances in the local file system during the database initialization process.
Once a context is created in the dictionary, context-specific databases can be populated using the context name.

A new feature extractor can be integrated into GBD by registering a feature extractor function in a dictionary data structure that is initialized in the \textsf{gbd\_init.feature\_extractors} module.
The dictionary is indexed by the \emph{name} of the feature extractor and contains a \emph{list of the provided features}, which are tuples of feature names and default values. 
It also contains a pointer to the feature extractor \emph{function} and a list of \emph{contexts} in which it can be used.
Registered feature extractors are automatically accessible by their names as subcommands of \textsf{gbd init}.

Instance transformers are managed in a dictionary data structure that is initialized in the \textsf{gbd\_init.instance\_transformers} module.
An instance transformer function takes an instance in a \emph{source context} and transforms it to a new instance in the \emph{target context}.
The dictionary is indexed by the \emph{name} of the instance transformer and, similarly to feature extractors, contains a \emph{list of the provided features}.
Additionally, the dictionary contains a pointer to the instance transformer \emph{function itself}, as well as a pointer to a function that generates the \emph{name of the benchmark instance} to be created.
Registered instance transformers are automatically accessible by their names as subcommands of \textsf{gbd transform}.

\section{Related and Future Work}\label{sec:outlook}

\textsf{SatLib} was the first public collection of benchmark instances in the SAT problem domain~\cite{Hoos:2000:SatLib}, and similar collections have been created for other problem domains such as MaxSAT~\cite{maxsatlib}, Quantified Boolean Formulas (QBF)~\cite{qbflib}, SAT Modulo Theories (SMT)~\cite{smtlib}, and Mixed Integer Programming (MIP)~\cite{miplib}.
\textsf{SatEx} is the first web-based framework for reproducible execution and evaluation of SAT solver experiments~\cite{Simon:2001:SatEx}, followed by the \textsf{EDACC} framework~\cite{Balint:2010:EDACC}.
Instance features were used to predict the fastest solver for an instance~\cite{Hoos:2008:SATzilla} and to reduce redundancy in experiments~\cite{Manthey:2016:Structure}.
\textsf{Aslib} is a library of data sets for training and evaluating solver prediction models~\cite{Hoos:2016:Aslib}.
Meta-features such as the instance domain have been used to study specialized heuristic configurations~\cite{Elffers:2018:Heuristics}.

GBD's data model, which is centered around instance identification functions, enables the data-driven study of NP-hard problem domains in an unprecedentedly sustainable manner.
Future work includes the integration of more problem domains, more feature extractors, and more instance transformers.
We also plan to improve the configuration capabilities of the data sources, including the ability to easily switch between different configurations, and to improve the extensibility of GBD by making it a matter of changing a configuration file whenever new problem domains, instance formats, feature extractors, and instance transformers are added.
We also plan to improve the web interface to make it easier to use and more informative, and to provide a benchmark submission system.
Finally, we plan to automate the feature extraction process to instantly provide feature databases for new benchmark instances, and add support for instance generators.

The purpose of coupling and providing instance data is to enable more complex data-driven analyses of the instance spaces of hard algorithmic problems, thus accelerating the growing use of explainable artificial intelligence methods.
Future work in this area includes the application of machine-assisted hypothesis generation and testing, and the development of new methods for exploring, analyzing, and explaining algorithmic datasets.

\bibliography{main.lipics}

@article{SC2020:AIJ,
  author	= {Nils Froleyks and Marijn Heule and Ashlin Iser and Matti Järvisalo and Martin Suda},
  title		= {{SAT} {C}ompetition 2020},
  journal	= {Artif. Intel.},
  year		= {2021},
  doi		= {10.1016/j.artint.2021.103572},
  durl		= {https://www.sciencedirect.com/science/article/pii/S0004370221001235}
}

@inproceedings{Iser:SC2023Bench,
  title		= {Benchmark {C}ompilation for {SAT} {C}ompetition 2023},
  author	= {Ashlin Iser},
  booktitle	= {Proc. of SAT Comp.~2023: Solver, Benchmark and Proof Checker Descriptions},
  year		= {2023},
  dcrossref	= {SC2023},
  url = {http://hdl.handle.net/10138/563824}
}

@inproceedings{Elffers:2018:Heuristics,
  author	= {Jan Elffers and Jes{\'{u}}s Gir{\'{a}}ldez{-}Cru and Stephan Gocht and Jakob Nordstr{\"{o}}m and Laurent Simon},
  title		= {Seeking {P}ractical {CDCL} {I}nsights from {T}heoretical {SAT} {B}enchmarks},
  booktitle	= {Intl. Joint Conf. on Artif. Intel., {IJCAI}},
  pages		= {1300--1308},
  year		= {2018},
  doi     = {10.24963/ijcai.2018/181}
}

@inproceedings{Haberlandt:2023:SBVA,
  author	= {Andrew Haberlandt and Harrison Green and Marijn J. H. Heule},
  title		= {Effective {A}uxiliary {V}ariables via {S}tructured {R}eencoding},
  booktitle	= {Intl. Conf. on Theo. and Appl. of Satisf. Test., {SAT}},
  pages		= {1--19},
  year		= {2023},
  doi		  = {10.4230/LIPICS.SAT.2023.11}
}

@phdthesis{Schreiber:2023:Diss,
  author	= {Schreiber, Dominik Pascal},
  year		= {2023},
  title		= {Scalable {SAT} {S}olving and its {A}pplication},
  doi		  = {10.5445/IR/1000165224},
  school	= {Karlsruhe Institute of Technology, {KIT}}
}

@inproceedings{Iser:2018:GBD,
  author	= {Ashlin Iser and Carsten Sinz},
  title		= {A {P}roblem {M}etadata {L}ibrary for {R}esearch in {SAT}},
  booktitle	= {Proceedings of Pragmatics of {SAT}, {POS}},
  pages		= {144--152},
  year		= {2018},
  doi     = {10.29007/gdbb}
}

@inproceedings{Bach:2022:kPortfolios,
  author	= {Jakob Bach and Ashlin Iser and Klemens B{\"{o}}hm},
  title		= {{A} {C}omprehensive {S}tudy of k-{P}ortfolios of {R}ecent {SAT} {S}olvers},
  booktitle	= {Intl. Conf. on Theo. and Appl. of Satisf. Test., {SAT}},
  pages		= {1--18},
  year		= {2022},
  doi     = {10.4230/LIPIcs.SAT.2022.2}
}

@article{Balint:2010:EDACC,
  author	= {Adrian Balint and Daniel Gall and Gregor Kapler and Robert Retz},
  title		= {{E}xperiment {D}esign and {A}dministration for {C}omputer {C}lusters for {SAT} {S}olvers, {EDACC}},
  journal	= {J. Satisf. Bool. Model. Comput.},
  pages		= {77--82},
  year		= {2010},
  durl		= {https://doi.org/10.3233/sat190078},
  doi		  = {10.3233/SAT190078}
}

@article{Simon:2001:SatEx,
  author	= {Laurent Simon and Philippe Chatalic},
  title		= {Sat{E}x: {A} {W}eb-based {F}ramework for {SAT} {E}xperimentation},
  journal	= {Electron. Notes Discrete Math.},
  pages		= {129--149},
  year		= {2001},
  url = {https://doi.org/10.1016/S1571-0653(04)00318-X}
}

@article{Hoos:2000:SatLib,
  author	= {Hoos, Holger and Stützle, Thomas},
  title		= {{SATLIB}: {A}n {O}nline {R}esource for {R}esearch on {SAT}},
  year		= {2000},
  pages		= {283-292},
  journal	= {SAT 2000},
  url     = {https://api.semanticscholar.org/CorpusID:17486963}
}

@article{Hoos:2016:Aslib,
  author	= {Bernd Bischl and Pascal Kerschke and Lars Kotthoff and others},
  title		= {{AS}lib: {A} {B}enchmark {L}ibrary for {A}lgorithm {S}election},
  journal	= {Artif. Intel.},
  pages		= {41--58},
  year		= {2016},
  doi = {https://doi.org/10.1016/j.artint.2016.04.003}
}

@inproceedings{Manthey:2016:Structure,
  author	= {Möhle, Sibylle and Manthey, Norbert},
  title		= {{B}etter {E}valuations by {A}nalyzing {B}enchmark {S}tructure},
  booktitle	= {Pragmatics of {SAT} workshop, {POS}},
  year		= {2016},
  url     = {http://www.pragmaticsofsat.org/2016/reg/POS-16_paper_4.pdf}
}

@book{Papadimitriou:1994:Complexity,
  author	= {Christos H. Papadimitriou},
  title		= {{C}omputational {C}omplexity},
  year		= {1994},
  publisher    = {Addison-Wesley}
}

@misc{qbflib,
  author	= {E. Giunchiglia and M. Narizzano and L. Pulina and A. Tacchella},
  title		= {Quantified {B}oolean {F}ormulas {L}ibrary, {QBFLIB}},
  url 		= {www.qbflib.org},
  year		= {2005}
}

@misc{maxsatlib,
  author	= {Fahiem Bacchus},
  title		= {Max{SAT} {F}ormulas {L}ibrary},
  url 		= {http://www.cs.toronto.edu/maxsat-lib/maxsat-instances/},
  year		= {2006}
}

@article{miplib,
  author	= {Gleixner, Ambros and Hendel, Gregor and Gamrath, Gerald and others},
  doi		  = {10.1007/s12532-020-00194-3},
  journal	= {Math. Program. Comput.},
  pages		= {443--490},
  title		= {{MIPLIB} 2017: {D}ata-driven {C}ompilation of the 6th {M}ixed-{I}nteger {P}rogramming {L}ibrary},
  year		= {2021}
}

@misc{smtlib,
  author =	 {Clark Barrett and Pascal Fontaine and Cesare Tinelli},
  title =	 {The {S}atisfiability {M}odulo {T}heories {L}ibrary, {SMT-LIB}},
  url = {www.smt-lib.org},
  year = 2016
}

@phdthesis{Fielding:2000:REST,
  author	= {Fielding, Roy Thomas},
  school	= {University of California},
  title		= {{REST:} {A}rchitectural {S}tyles and the {D}esign of {N}etwork-based {S}oftware {A}rchitectures},
  url		  = {http://www.ics.uci.edu/~fielding/pubs/dissertation/top.htm},
  year		= {2000}
}

@inproceedings{Mckinney:2010:Pandas,
  title		= {{D}ata {S}tructures for {S}tatistical {C}omputing in {P}ython},
  author	= {McKinney, Wes and others},
  booktitle	= {{P}ython in {S}cience},
  pages		= {51--56},
  year		= {2010},
  doi     = {10.25080/majora-92bf1922-00a}
}

@article{Hoos:2008:SATzilla,
  author = {Xu, Lin and Hutter, Frank and Hoos, Holger H. and Leyton-Brown, Kevin},
  title = {{SAT}zilla: {P}ortfolio-based {A}lgorithm {S}election for {SAT}},
  year = {2008},
  journal = {J. Artif. Intel. Res., JAIR},
  pages = {565--606},
  doi = {10.5555/1622673.1622687}
}

@misc{Iser_GBD_Tools_2023,
  author = {Iser, Ashlin and Jabs, Christoph},
  doi = {10.5281/zenodo.11093597},
  title = {{GBD Tools}},
  url = {https://github.com/Udopia/gbd},
  version = {4.8.5},
  year = {2024}
}

@article{Merkel:2014:Docker,
  title={Docker: lightweight linux containers for consistent development and deployment},
  author={Merkel, Dirk},
  journal={Linux J.},
  year={2014}
}

@article{Reese:2008:Nginx,
  author = {Reese, Will},
  title = {Nginx: the high-performance web server and reverse proxy},
  journal = {Linux J.},
  year = {2008},
}

\end{document}